\pdfoutput=1
\documentclass[letterpaper, final, conference]{ieeeconf}
\IEEEoverridecommandlockouts 
\usepackage{times}
\usepackage[utf8]{inputenc}

\usepackage{microtype}
\usepackage{graphicx}
\usepackage{epstopdf}
\graphicspath{{figures/}}
\DeclareGraphicsExtensions{.eps,.pdf,.jpeg,.png}

\usepackage{float}
\usepackage{tabularx}
\usepackage{booktabs}
\usepackage[caption=false]{subfig}
\usepackage{array}
\usepackage{multirow}
\newcolumntype{L}{>{$}l<{$}} 
  \newcolumntype{R}{>{$}r<{$}} 
\newcolumntype{C}{>{$}c<{$}} 
\newcolumntype{P}[1]{>{\centering\arraybackslash}p{#1}}
\let\labelindent\relax
\usepackage[inline]{enumitem}
\usepackage{xcolor}

\usepackage[hyphens]{url}
\usepackage[
  bookmarks=true,
]{hyperref}

\usepackage{algorithm}
\usepackage{algpseudocode}

\usepackage{notation}

\usepackage{amsthm}
\theoremstyle{definition}
\newtheorem{definition}{Definition}
\newtheorem*{definition*}{Definition}

\newtheorem{problem}{Problem}

\theoremstyle{plain}
\newtheorem{proposition}{Proposition}

\theoremstyle{remark}

\newtheorem*{remark}{Remark}
\newtheorem*{note}{Note}
\newtheorem{example}{Example}

\makeatletter
\renewcommand\paragraph{\@startsection{paragraph}{4}{\z@}%
  {-12\p@ \@plus -4\p@ \@minus -4\p@}%
  {-0.5em \@plus -0.22em \@minus -0.1em}%
  {\normalfont\normalsize\bfseries}}
\makeatother

\usepackage{ifdraft}
\usepackage{comment}
\usepackage[obeyFinal]{todonotes}

\newenvironment{todocomment}[1][]{
  \par\textcolor{red}{\bfseries To-Do\ifblank{#1}{}{ (#1)}:} \color{red}\ignorespaces%
}{
  \par
}

\ifoptionfinal{%
  \excludecomment{todocomment}
}{%
  \usepackage[columnwise,switch]{lineno}
  \linenumbers

}

\title{Motion Planning for Automata-based Objectives using Efficient
  Gradient-based Methods}

\author{%
  \authorblockN{%
    Anand~Balakrishnan,
    Merve~Atasever,
    Jyotirmoy V.~Deshmukh
  }
  \authorblockA{%
    University of Southern California, Los Angeles, California, USA\\
    Email: \{anandbal, atasever, jdeshmuk\}@usc.edu%
  }
  \thanks{This work was partially supported by the National Science Foundation
    through the following grants: CAREER award (SHF-2048094), CNS-1932620,
    CNS-2039087, FMitF-1837131, CCF-SHF-1932620, funding by Toyota R\&D and Siemens Corporate Research through the USC Center for Autonomy and AI, an Amazon Faculty Research Award, and the Airbus Institute for Engineering Research.}
}

\begin{document}

\maketitle

\begin{abstract}

  In recent years, there has been increasing interest in using formal
  methods-based techniques to safely achieve temporal tasks, such as timed
  sequence of goals, or patrolling objectives.
  Such tasks are often expressed in real-time logics such as Signal Temporal
  Logic (STL), whereby, the logical specification is encoded into an
  optimization problem.
  Such approaches usually involve optimizing over the quantitative semantics, or
  robustness degree, of the logic over bounded horizons: the semantics can be
  encoded as mixed-integer linear constraints or into smooth approximations of
  the robustness degree.
  A major limitation of this approach is that it faces scalability challenges
  with respect to temporal complexity: for example, encoding long-term tasks
  requires storing the entire history of the system.
  In this paper, we present a quantitative generalization of such tasks in the
  form of symbolic automata objectives.
  Specifically, we show that symbolic automata can be expressed as matrix
  operators that lend themselves to automatic differentiation, allowing for the
  use of off-the-shelf gradient-based optimizers.
  We show how this helps solve the need to store arbitrarily long system
  trajectories, while efficiently leveraging the task structure encoded in the
  automaton.

\end{abstract}

\IEEEpeerreviewmaketitle


\section{Introduction}


For autonomous robots operating in highly uncertain or dynamic environments,
motion planning can be challenging \cite{lavalle2006planning,kavraki2016motion}.
A popular class of motion planning algorithms that address such environments is
\emph{model predictive control}, which recomputes short-term plans in real-time
to adapt to changes in the environment or stochasticity in the environment
dynamics \cite{richards2005robust,camacho2013model}.
In this design paradigm, the system designer creates a predictive model of the
behavior of the robot and its operating environment that is used at runtime to
predict future states based on some finite sequence of control actions.
The system uses this \emph{model predictive} approach along with an online
optimizer to fine the optimal sequence of control actions that minimize some
user-specified cost function on the predicted trajectory.
The system then applies the first control action, and then replans the sequence
for the newly observed state.
This paradigm is also referred to as \emph{receding horizon} planning.

In most modern motion planning approaches, the optimization problem is directly
or indirectly reduced to solving quadratic cost functions over system states and
actions, with the assumption that the desired system behaviors are the ones that
minimize such costs \cite{katayama2023model,nguyen2021model}.
While this is adequate for task objectives such as tracking a set of way-points,
or minimizing the energy consumed by the robot, some objectives
require the robot to exhibit complex spatio-temporal behavior
\cite{kress-gazit2009temporallogicbased,finucane2010ltlmop,menghi2021specification}.
There is growing body of literature to specify such tasks using formalisms such
as Linear Temporal Logic (LTL) \cite{pnueli1977temporal} and Signal Temporal
Logic (STL) \cite{maler2004monitoring} and the use of these logics for robot
motion planning.

Broadly speaking, there have been two high-level directions one can take with
motion planning with temporal logic specifications \cite{belta2019formal}:
\begin{enumerate*}
  \item by translating the specification into an automaton, and decomposing the
        planning problem over it; or
  \item by using the \emph{quantitative semantics} of the logic to directly
        optimize for satisfaction or robustness of the system w.r.t.\ the
        specification.
\end{enumerate*}

The techniques presented in \cite{fainekos2009temporal, finucane2010ltlmop,
  bhatia2010samplingbased,wongpiromsarn2012receding, lahijanian2016iterative}
represent the former of the above approaches.
Here, the frameworks deconstruct complex, temporal motion planning over the
transitions in an automaton representing the temporal specification.
Specifically, for each location in the automaton, a corresponding set is
computed in the state space of the system, and the motion plan is computed for
each pair of such connected sets.
These approaches suffer from the lack of scalability, as the size of the
automata can increase exponentially to that of the temporal logic specification.

On the other hand, several proposed works in recent literature directly optimize
over the semantics of the temporal logic.
Of particular relevance to this paper is STL which allows defining a
\emph{robust satisfaction value} or \emph{robustness} that approximates the
distance of a given trajectory from the set of trajectories satisfying the
formula \cite{fainekos2009robustness,donze2010robust,donze2013efficient}.
The robustness metric is leveraged to encode the motion planning problem for
linear (and piecewise-linear) dynamical systems as a mixed integer linear
program \cite{farahani2015robust,raman2014model} or by gradient-based
optimization of the smooth approximation of robustness
\cite{haghighi2019control,pant2017smooth}.
A key limitation of these approaches are that they are not applicable to general
nonlinear models, or suffer from intractability with increasing prediction
horizon and formula complexity.



\paragraph*{Our Contributions}
In this paper, we propose a framework to bridge the gap between automata-based
techniques and robustness-based optimization.
Specifically, we define a \emph{matrix operator} on symbolic automata that
\begin{enumerate}
  \item translates symbolic automata to weighted matrices given a system state;
        and
  \item builds on abstract matrix operations to enable automatic differentiation.
\end{enumerate}
This matrix operator can be used along with off-the-shelf gradient-based
optimization pipelines to compute end-to-end motion plans for complex temporal
specifications.
We will demonstrate the efficacy of our proposed framework in various motion
planning scenarios by comparing against similar tools.
We also explore how various notions of \emph{robustness} can be achieved within
the same pipeline, and how they manifest in concrete tasks.

\subsection*{Related Work}

In automata-based methods, temporal logic specifications are usually translated
to finite state automata (for finite-length behavior) or \(\omega\)-regular
automata (for infinite behavior)~\cite{bartocci2018specificationbased}.
The control problem is then reduced to a graph game on the product of the
specification automaton and a finite model of the system, such as a transition
system or a Markov Decision Process~\cite{baier2008principles,wongpiromsarn2012receding}; or as a hierarchical control problem on
a hybrid
system~\cite{kress-gazit2009temporallogicbased,fainekos2009temporal,smith2011optimal,lindemann2020efficient}.
The main limitation of automata-based methods is the prohibitive computational
complexity, due to the exponential blow-up of finite model abstractions for
infinite systems.

The main thrust of optimization-based methods are temporal
logics with semantics defined over finite-length signals, namely Signal Temporal
Logic \cite{maler2004monitoring} and Metric Temporal Logic
\cite{koymans1990specifying}.
Specifically, many optimization-based approaches exploit the quantitative
semantics defined for these logics presented in \cite{fainekos2009robustness}
and \cite{donze2010robust}.
These \emph{robustness} metrics are used in optimization pipelines to assign
costs that penalize deviation from specified system behavior.
For example, \cite{raman2014model} and \cite{sadraddini2019formal} both present
methods to translate STL formulas into mixed-integer constraints for convex
optimization, based on prior work for Linear Temporal Logic
\cite{wolff2014optimizationbased}.
Likewise, \cite{pant2017smooth} and \cite{haghighi2019control} present smooth
approximations of STL quantitative semantics to enable gradient-based
optimization directly over the STL formula.
Unlike automata-based approaches, such methods require entire signal histories
to compute a robustness score, and the \(\min\) and \(\max\) operations in the
quantitative semantics, can lead to various local optima due to vanishing
gradients and non-linearity.

Recent developments in \emph{automatic differentiation} algorithms
\cite{baydin2018automatic} has enabled a boom in the use of gradient-based
optimization methods, via off-the-shelf libraries like PyTorch
\cite{paszke2019pytorch}, and JAX \cite{frostig2018compiling}.
The works in \cite{haghighi2019control} and \cite{pant2017smooth} leverage this
in when defining their smooth approximations of robustness.
Similarly, the library presented in \cite{leung2021backpropagation} builds on
PyTorch to define the quantitative semantics of STL as \emph{compute graphs}.
Such gradient-based methods have allowed for the evaluation of
temporal logic quantitative semantics to scale well by leveraging the compute
pipeline that such libraries build on, but still suffer from dependence on
history.


\section{Preliminaries}

In this section, we will define some notation and background for our proposed
method.
Through this paper, we will use $S^n$ to denote the state space of our system,
where $S \subseteq \Re$.
Then, we can define predicates on \(S^n\) as Boolean expressions with the
recursive grammar:
\begin{equation}
  \label{eq:predicate-syntax}
  \varphi := \top \mid \bot \mid \mu(x) \geq 0
  \mid \varphi \land \varphi \mid \varphi \lor \varphi,
\end{equation}
where
\begin{itemize}
  \item \(\top\) and \(\bot\) refer to \emph{true} and \emph{false} values
        respectively;
  \item \(\mu: S^n \to \Re \) is a scalar, differentiable function; and
  \item \(\varphi \land \varphi\), and \(\varphi \lor \varphi\) refer to Boolean conjunction (\emph{and}), and disjunction (\emph{or}) operations respectively.
\end{itemize}
Let \(\Phi\) denote the set of all such predicates over \(S^n\).
For some \(s\in S^n\) and \(\varphi \in \Phi\), we say that \(s\) \emph{models}
\(\varphi\) (denoted \(s \models \varphi\)) if \(s\) satisfies the Boolean
predicate \(\varphi\).

\begin{remark}
  The above syntax is similar to that of Signal Temporal
  Logic~\cite{maler2004monitoring} without the temporal operators.
  Moreover, subsequent definitions will leverage this to define some concepts
  introduced in \cite{jaksic2018algebraic}.
\end{remark}

\begin{figure}
  \centering
  \includegraphics[width=0.366\textwidth]{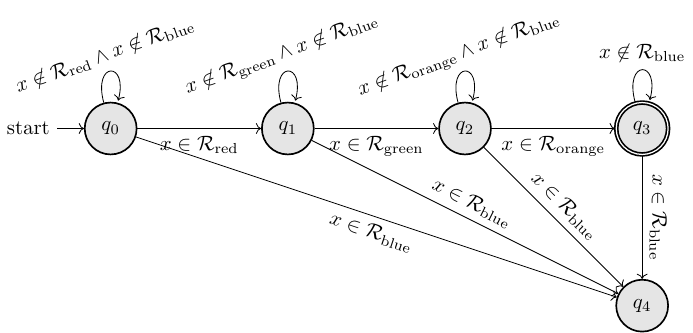}
  \caption{%
    Example of a symbolic automaton describing the specification ``move to
    region \(\Rc_{\text{red}}\), then region \(\Rc_{\text{green}}\), and then to
    region \(\Rc_{\text{orange}}\) in order while always avoiding region
    \(\Rc_{\text{blue}}\).
    ''
    An \emph{accepting} run in the automaton is a sequence of states \(\vect{x}
    \in \Sigma^*\) that moves the automaton location from
    the initial location \(q_0\) to the accepting location \(q_3\).%
  }\label{fig:sa-example}
\end{figure}

\begin{figure}
  \centering
  \includegraphics[width=0.5\columnwidth]{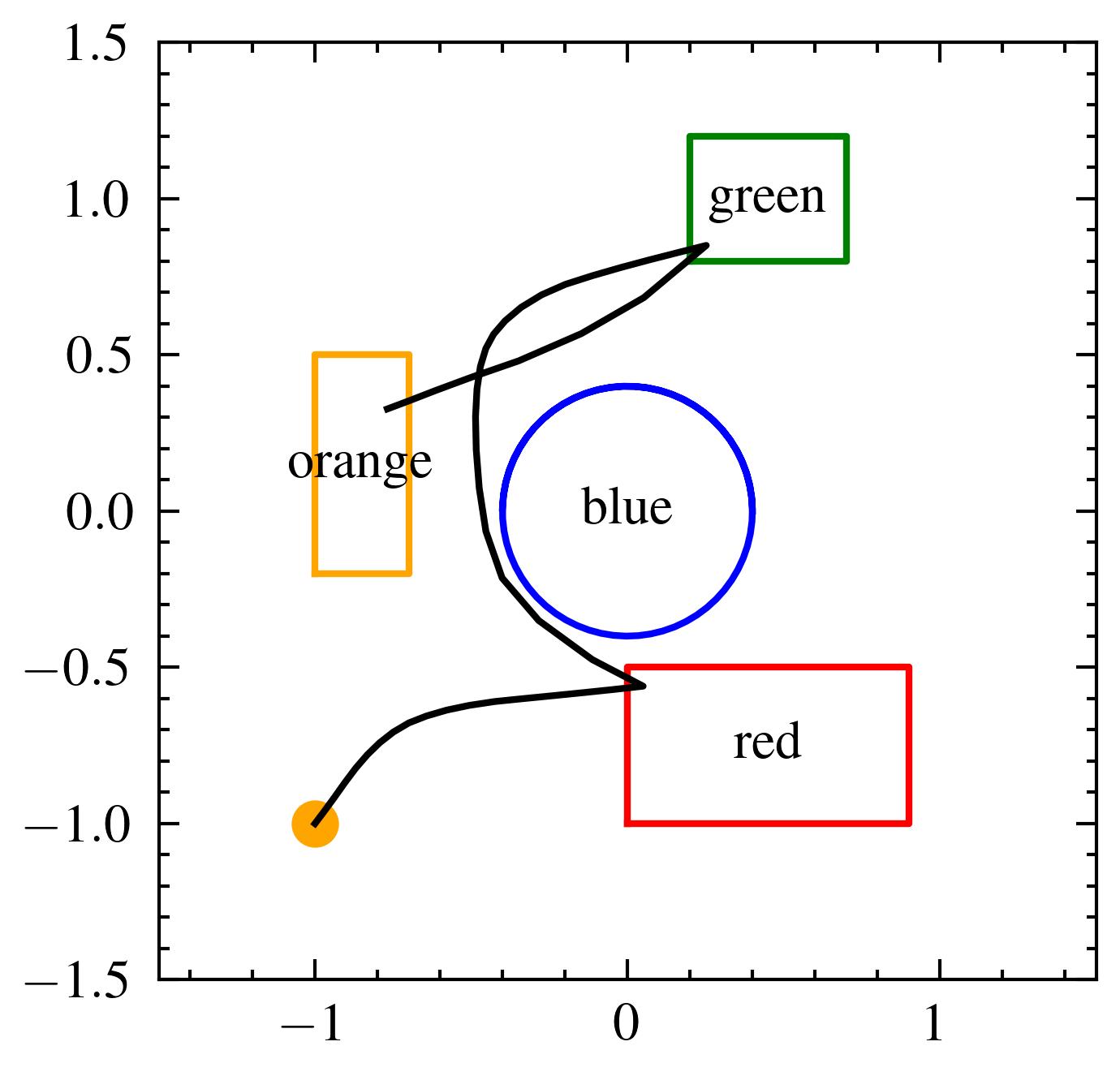}
  \caption{%
    An example trajectory in a 2D workspace that satisfies the specification
    in \autoref{fig:sa-example}.
    Here, the state \(x\) is a vector in \(\Re^2\), and the trajectory starts
    at a point \((-1, -1)\) and completes the specified task.
  }\label{fig:accepting-trajectory}
\end{figure}

\begin{definition}[Symbolic Automata \cite{dantoni2017power}]
  A symbolic automaton is a tuple \(\Ac = \left( \Sigma, Q, Q_0, Q_F, \Delta
  \right)\) where
  \begin{itemize}
    \item \(\Sigma\) is an input alphabet;
    \item \(Q\) is a set of \emph{locations} in the automaton, with \(Q_0\) and
          \(Q_F\) denoting the initial and final (or accepting) locations
          respectively; and
    \item \(\Delta: Q \times Q \to \Phi\) denotes a mapping from transitions in
          the automaton to a Boolean predicate expression.
  \end{itemize}
\end{definition}

In this paper, we restrict ourselves to automata that accept \emph{regular
  languages}, i.e., automata that are defined on finite length sequences of
elements in the input alphabet.
We use \(\Sigma^*\) to refer to the set of all finite length
sequence of elements \(s \in \Sigma\).
Given a sequence of input alphabets, \(\xi = (s_0, s_1, \ldots, s_l) \in
\Sigma^*\), a \emph{run} in the symbolic automaton \(\Ac\) is a sequence of
locations \((q_0, q_1, \ldots, q_{l+1})\) such that \(s_i \models \Delta(q_i, q_{i+1})\).
We use \(q_i \xrightarrow{s_i} q_{i+1}\) to denote such a valid transition in
the automaton, and \(\Run_\Ac(\xi)\) to denote the set of runs induced in \(\Ac\) by
\(\xi \in \Sigma^*\).
If single trace \(\xi\) can generate multiple runs in \(\Ac\), we say that the
automaton is non-deterministic.
A run is \emph{accepting} in the automaton if \(q_l \in Q_f\) for any induced
run \((q_0, q_1,
\ldots, q_l) \in \Run_\Ac(\xi)\), and we denote this by \(\xi
\models \Ac\).

\begin{remark}
  Note that a run \(\Run_\Ac(\xi)\) is \emph{rejecting} in \(\Ac\) if the last
  location in the run is not in \(Q_F\).
  Thus, for a finite state automaton, the logical negation of the specification
  can be obtained by checking if a run does not end in \(Q_F\), i.e., \(Q_F' = Q
  \setminus Q_F\).
\end{remark}

\begin{example}
  In
  \autoref{fig:sa-example}
  and
  \autoref{fig:accepting-trajectory},
  we see an example of a symbolic automaton with a sequential specification, and
  a trajectory that satisfies it.
\end{example}

\begin{definition}[Semiring~\cite{mohri2002semiring}]
  A tuple, \(\Ke = \Tuple{K, \oplus, \otimes, \tilde{0}, \tilde{1}}\) is
  a \emph{semiring} with the underlying set \(K\) if \(\left(K, \oplus,
  \tilde{0}\right)\) is a commutative monoid with identity \(\tilde{0}\);
  \(\left( K, \otimes, \tilde{1} \right)\) is a monoid with identity element
  \(\tilde{1}\); \(\otimes\) distributes over \(\oplus\); and \(\tilde{0}\) is
  an annihilator for \(\otimes\) (for all \(k \in K, k \otimes \tilde{0} =
  \tilde{0} \otimes k = \tilde{0}\)).
\end{definition}

\begin{definition}[Symbolic Weighted
    Automata~\cite{droste2009handbook,dantoni2017power,jaksic2018algebraic}]
  Given a symbolic automaton \(\Ac\) and a semiring \(\Ke = \left( K, \oplus,
  \otimes, \tilde{0}, \tilde{1} \right)\), a symbolic \emph{weighted} automaton
  over a semiring \(\Ke\) is a tuple  \(\left( \Ac, w \right)\), where
  \(w: Q \times \Sigma \times Q \to K\) is a weighting function.
\end{definition}

We define the weight of a sequence of inputs \(\xi \in \Sigma^*\) in an symbolic
weighted automaton \(\Ac\) as the mapping \(w_\Ac: \Sigma^* \to K\), where
\begin{displaymath}
  w_\Ac(\xi) = \bigoplus_{r \in \Run_\Ac(\xi)} \bigotimes_{i=0}^{\abs{\xi}} w(r_i, \xi_i, r_{i+1})
\end{displaymath}

\begin{remark}
  Note that when using \(\Ke = \Tuple{\Ne \cup \{\infty\}, \min, +, 0,
    \infty}\), this definition of \(w\) is equivalent to that of the standard
  shortest path in a directed graph.
\end{remark}

\begin{example}[Weights of a path]
  Looking at the same automata as in \autoref{fig:sa-example}, let us define
  a weighting function as
  \begin{displaymath}
    w(q, x, q') =
    \begin{cases}
      0, & ~\text{if}~q' = q_4 \\
      1, & ~\text{otherwise.}
    \end{cases}
  \end{displaymath}
  This weighting function assigns \(0\) weight if the input sequence enters the
  rejecting sink state \(q_4\).
  Under different semirings, we can see different effects of \(w_\Ac\) as
  follows (noting that \(\abs{\Run_\Ac(\xi)} = 1\) for any \(\xi \in \Sigma^*\)
  as \(\Ac\) is \emph{deterministic}):
  \begin{itemize}
    \item Boolean Semiring \(\Ke = \left( \left\{ 0, 1 \right\}, \lor, \land, 0,
          1 \right)\):
          Here, any input sequence \(\xi\) that does not induce a run that
          enters \(q_4\) will have \(w_\Ac(\xi) = 1\) as all weights will be
          \(1\) (which is equivalent to the Boolean true value \(\top\)).
          Otherwise, \(w_\Ac(\xi) = 0\), as even a single \(0\) weight
          transition will render the conjunction \(\otimes \equiv \land\) false.

    \item \((\min, \max)\) Semiring \(\Ke = (\{0,1\}, \max, \min, 0, 1)\): This
          is semantically equivalent to the Boolean semiring, and thus produced
          identical results.

    \item \((\min, +)\) Semiring \(\Ke = (\Ne \cup \{\infty\}, \min, +,
          0,\infty)\): Referred to as a \emph{tropical} semiring, under this
          semiring, \(w_\Ac(\xi)\) outputs the length of the input sequence
          until it first reaches the rejecting sink \(q_4\) or the length of
          the input sequence itself \(\abs{\xi}\).

  \end{itemize}
\end{example}

\paragraph*{Matrix Semirings}

If \(m\) is a positive integer and \(K\) is a semiring, then the set of \(m
\times m\) matrices with entries in \(K\), denoted \(K^{m \times m}\), is also
a semiring \cite{golan2003semirings}.
Specifically, for matrices \(A, B, C \in K^{m \times m}\), we can define the
semiring operation as follows:
\begin{itemize}
  \item Addition \(A \oplus B = C\) is defined as element-wise addition \(C_{ij}
        = A_{ij} \oplus B_{ij}\);
  \item The additive identity matrix is, intuitively, the \(m \times m\) matrix
        with all entries \(\tilde{0}\);
  \item Multiplication \(AB = C\) is defined similar to matrix multiplication as
        \(C_{ij} = \bigoplus_{k=0}^{m - 1} A_{ik} \otimes B_{kj}\); and
  \item The multiplicative identity matrix (or simply, identity matrix) is
        similar to the usual: an \(m \times m\) matrix with all diagonal entries
        equal to \(\tilde{1}\), and the rest are \(\tilde{0}\).
\end{itemize}

\begin{remark}
  Note that the above translates to the usual matrix multiplication in linear algebra for the semiring of reals \(\left( \Re, +, \times, 0, 1\right)\).
\end{remark}

From the above definition of matrix semiring arithmetic, one can derive the
vector dot product, the vector-matrix product, and various other concepts from
linear algebra that show direct equivalences under the abstract algebraic
framework~\cite{golan2003semirings}.
In the rest of this paper, we will use the standard notation for matrix
multiplication (\(C =AB\)), vector dot product (\(x_3 = x_1 \cdot x_2 = x_1 x_2\)), and
vector-matrix multiplication (\(x_2 = x_1^T A\)), but the operations are defined
on semirings unless otherwise specified.


\section{Motion Planning with Automaton Matrix Operators}

For optimization-based motion planning, we concern ourselves with two general
motion planning problems given temporal specifications:

\begin{problem}[Open-Loop Motion Planning]
Given a discrete-time, dynamical system
\begin{equation}
  \label{eq:dt-dynamics}
  x_{t+1} = f(x_t, u_t),
\end{equation}
where \(x_{t}, x_{t+1} \in S^n\), \(u_t \in \Uc \subseteq \Re^m\), and \(f:
S^n \times \Uc \to S^n\) is a (piecewise) differentiable function.
For a task automaton \(\Ac\), a planning horizon \(H \in \Ne\), and some
initial state \(x_0 \in \Sigma = S^n \), compute a control plan \(\vect{u}^* = \left(
u_0^*,\ldots,u_{H-1}^*\right)\), such that, for \(t \in 0, \ldots, H-1\):
\begin{equation}
  \begin{array}{rcl}
    \vect{u}^* & = & \argmin_{\vect{u}} \norm{\vect{u}}          \\
    \text{~s.t.~}
               &   & \left( x_0, \ldots, x_H \right) \models \Ac \\
               &   & x_{t+1} = f(x_t, u_t).
  \end{array}
\end{equation}
\end{problem}

\begin{problem}[Closed-loop Control]
Given a discrete-time, dynamical system like in
\autoref{eq:dt-dynamics} and a task automaton \(\Ac\), derive a feed-back
control law
\begin{equation}
  \begin{array}{rcl}
    u_t^* & = & \pi(x_t)                       \\
    \text{~s.t.~}
          &   & (x_0, x_1, \ldots) \models \Ac \\
          &   & x_{t+1} = f(x_t, u_t^*)
  \end{array}
\end{equation}
\end{problem}

In this section, we will define the \emph{automaton matrix operator} \(\Ae\) for
a symbolic automaton \(\Ac\).
The matrix operator for an automaton \(\Ac\) is a mapping \(\Ae: \Sigma \to
K^{\abs{Q} \times \abs{Q}}\) that, for each input element in the alphabet \(x \in
\Sigma\) maps to a \(\abs{Q}\times \abs{Q}\) matrix with entries in the set
\(K\) with some semiring associated with it.
The goal of this mapping is to capture the structure of the automaton while
leveraging matrix semiring algebra~\cite{golan2003semirings} to compute weights
of system trajectories and, consequently, reason about the acceptance of the
given system trace in the automaton.

Before we formally define \(\Ae\), we will define a few weighting functions for
it:
\begin{definition}[Generalized Weights]%
  For a given predicate \(\varphi \in \Phi\), some value \(x \in S^n\)
  and a semiring \(\Ke = \Tuple{K \subseteq \Re, \oplus, \otimes, \tilde{0},
    \tilde{1}}\),
  let \(\lambda: S^n \times \Phi \to K\) be recursively defined as follows
  \begin{equation}
    \renewcommand{\arraystretch}{1.5}
    \begin{array}{lcl}
      \lambda(x, \top) = \tilde{1},         &   & \lambda(x, \bot) = \tilde{0}                        \\
      \lambda(x, \mu(x) \geq 0)             & = &
      \begin{cases}
        \tilde{0} & ~\text{if}~ \mu(x) \geq 0 \\
        \mu(x)    & ~\text{if}~ \mu(x) < 0    \\
      \end{cases}
      \\
      \lambda(x, \varphi_1 \land \varphi_2) & = & \lambda(x, \varphi_1) \otimes \lambda(x, \varphi_2) \\
      \lambda(x, \varphi_1 \lor \varphi_2)  & = & \lambda(x, \varphi_1) \oplus
      \lambda(x, \varphi_2).
    \end{array}
  \end{equation}

  Let \(\alpha, \beta \in K^{\abs{Q}}\) be the initial and final weights
  respectively for \(\Ac\) such that:
  \begin{equation}
    \begin{array}{cc}
      \alpha_i =
      \begin{cases}
        \tilde{1}, & ~\text{if}~ q_i \in Q_0 \\
        \tilde{0}, & \text{~otherwise,}
      \end{cases}
       &
      \beta_i =
      \begin{cases}
        \tilde{1}, & ~\text{if}~ q_i \in Q_F \\
        \tilde{0}, & \text{~otherwise.}
      \end{cases}
    \end{array}
  \end{equation}
\end{definition}

In the above definitions, \(\lambda\) corresponds to a symbolic weighting
function that generates a weight in \(K\) for each concrete input \(x \in S^n\);
and
\(\alpha\) and \(\beta\) correspond to the initial and final locations in the
automaton respectively.

\begin{definition}[Automaton Matrix Operator]%
  For a given weighted automaton \(\left( \Ac, w \right)\), where \(w(q, s, q')
  = \lambda\of*{s, \Delta\of*{q, q'}}\), let \(\Ae: S^n \to K^{\abs{Q} \times
      \abs{Q}}\) be a matrix semiring operator over the semiring \(\Ke = \Tuple{K
    \subseteq \Re, \oplus, \times, \tilde{0}, \tilde{1}} \) such that:
  \begin{equation}
    \Ae(s)_{ij} = \lambda(s, \Delta(q_i,q_j))
  \end{equation}
\end{definition}

Note that the matrix operator \(\Ae(x)\) defines a \emph{weighted transition
  matrix} for the automaton where each entry \(\Ae(x)_{ij}\) determines the cost of
moving from location \(q_i\) to \(q_j\) when seeing the input \(x\).
Thus, given a previous weighted location vector \(q\) and an
input state \(x\), we can write the next weighted location vector as \(q' = q^T
\Ae(x)\).
Thus, we can define the weight of a state trajectory \(\xi = \left( x_0, x_1,
\ldots, x_l \right)\) from the set of initial locations \(Q_0\) (encoded in
\(\alpha\)) to any final location in \(Q_F\) (encoded in \(\beta\)) as follows:
\begin{equation}
  w_\Ac(\xi) = \alpha^T \Ae(x_0) \Ae(x_1) \ldots \Ae(x_l) \beta.
\end{equation}




By encoding the automaton as a matrix operator, and defining the semantics of
the weighted automaton trough matrix semirings, we are able to leverage
state-of-the-art automatic differentiation libraries built on matrix and array
operations.
Thus, \autoref{alg:open-loop} shows how we leverage this in a gradient-based
pipeline to solve the open-loop control problem.

From the above, we can solve the open-loop plan as the solution of a gradient-based
optimization problem using the procedure in \autoref{alg:open-loop}.
Specifically, the gradient \(\nabla_{\vect{u}} w_\Ac\) in line
\ref{algline:gradient-weight} can be symbolically computed using off-the-shelf
algorithms.
\autoref{alg:mpc} shows how we can use the open-loop algorithm as a subroutine
in computing a receding-horizon control law for satisfying \(\Ac\) by
\emph{memoizing} the current weight in \(\Ac\) at time \(t\) in the vector
\(q_t\), as seen in line \ref{algline:memoizing}, as in model predictive control
(MPC).
However, in general, motion planning is NP-Hard and gradient-based methods may
not guarantee convergence to an optimal solution.
They require either sufficiently good initial guesses or can be used in
combination with sampling-based methods like the cross-entropy
method~\cite{bharadhwaj2020modelpredictive}.

\begin{algorithm}
  \caption{Gradient-based optimization with automaton matrix operator.}%
  \label{alg:open-loop}
  \begin{algorithmic}[1]
    \Procedure{Open-Loop$_{\Ac,\Ke}$}{$x_{\text{init}}, q_{\text{init}},H, \gamma,
        k$}
    \Statex {\(x_{\text{init}} \in S^n\): initial system state}
    \Statex {\(q_{\text{init}} \in K^{\abs{Q}}\): an initial weight configuration}
    \Statex {\(H \in \Ne\): planning horizon}
    \Statex {\(\gamma > 0\): learning rate for gradient descent}
    \Statex {\(k \in \Ne\): number of optimization epochs}
    \State Zero-initialize \(\vect{u} = \left( u_0, \ldots, u_{H-1}\right)\)
    \State \(x_0 \gets x_{\text{init}}\)
    \For{\(1,\ldots,k\) epochs}
    \State \(\xi = \left( x_0, x_1, \ldots, x_H \right)\) from \autoref{eq:dt-dynamics}.
    \State Compute \(\Ae(x_i)\) for \(i \in 0, \ldots, H\).
    \State \(w_\Ac(\xi) = q_{\text{init}}^T \Ae(x_0) \Ae(x_1) \ldots \Ae(x_H) \beta\).
    \State \(\vect{u} \gets \vect{u} + \gamma \nabla_{\vect{u}}
    w_\Ac(\xi)\).\label{algline:gradient-weight}

    \EndFor
    \State\Return \(\vect{u}\)
    \EndProcedure%
  \end{algorithmic}
\end{algorithm}

\begin{algorithm}
  \caption{Gradient-based MPC with automaton matrix operator.}%
  \label{alg:mpc}
  \begin{algorithmic}[1]
    \Procedure{MPC$_{\Ac,\Ke}$}{$x_{\text{init}},H, \gamma, k$}
    \Statex {\(x_{\text{init}} \in S^n\): initial system state}
    \Statex {\(H \in \Ne\): planning horizon}
    \Statex {\(\gamma > 0\): learning rate for gradient descent}
    \Statex {\(k \in \Ne\): number of optimization epochs}
    \State \(x_0 \gets x_{\text{init}}\)
    \State \(q_0 \gets \alpha\)
    \For{\(t = 0,1,\ldots\)}
    \State \(\vect{u}^* \gets\) \Call{Open-Loop$_{\Ac,\Ke}$}{$x_t, q_t, H, \gamma, k$}
    \State Apply \(u_t^* \) (first item in \(\vect{u}^*\))
    \State Update \(x_{t+1}\) from environment
    \State \(q_{t+1} \gets q_{t}^T\Ae(x_t)\)\label{algline:memoizing}
    \EndFor
    \EndProcedure%
  \end{algorithmic}
\end{algorithm}

\section{Experimental Results}

To demonstrate the applicability of our method, we will show examples of
open-loop and closed-loop planning with some specifications relevant to
autonomous robots in general.
In all the experiments, we will compare against two other related works:
\begin{itemize}
  \item STLCG \cite{leung2021backpropagation}, where finite-length signal traces are
        evaluated over Signal Temporal Logic (STL) formula encoded as quantitative
        computation graphs in PyTorch~\cite{paszke2019pytorch}.
        This, along with our automaton operator method, will be used in
        a gradient-based optimization pipeline.
  \item Mixed integer program (MIP) encoding of STL robustness following
        \cite{raman2014model,belta2019formal} which will be optimized using
        off-the-shelf mixed integer convex program solvers \cite{kurtz2022mixedinteger}.
\end{itemize}

\begin{note}
  While the above frameworks are able to compute quantitative costs from logical
  specifications, one should note that the quantitative semantics of STL
  necessarily require the entire history of the system (unless restricted to a
  specific syntactic subset).
  Thus, this makes them generally unfeasible to use for \emph{all} MPC tasks,
  but we make best efforts to do so for our experiments.
  Specifically, closed-loop control for both, \(\varphi_1\) and \(\varphi_2\)
  in the experimental results is unfeasible using the above methods.
  This restriction is not present in our automaton-based approach.
\end{note}

In the case of open-loop experiments, we will report the total number of
optimizer iterations until a satisfying trajectory in the system is found
(denoted \(t^*\)).
Meanwhile, for closed-loop experiments, we will report the final robustness
\cite{donze2010robust} of the system trajectories with respect to the
corresponding STL task specifications(denoted \(\rho\)).
We report the results of our experiments in \autoref{tab:results-openloop}, with
the first column describing the specification under test.
\begin{remark}
  For the MILP solver, we report the number of simplex iterations performed by
  Gurobi~\cite{gurobi} for the open-loop problems.
  It should be noted that while Gurobi's simplex optimization iterations are
  about 10-20 times faster than a gradient-based solution on an Intel Core i7
  (1.80GHz) CPU machine with no graphics processor for computation purposes,
  the MILP solver is unfeasible for even relatively simple, long-horizon
  specifications.
\end{remark}

To aid our presentation, we will informally describe the syntax and semantics of
discrete-time STL (DT-STL).
In addition to the Boolean predicates defined in
\autoref{eq:predicate-syntax}, temporal logics add the following
temporal operators (relevant to our studies)
\begin{itemize}
  \item \(\Alw_{[a,b]} \varphi\), where \(a, b \in \Ne \cup \{ \infty \}\),
        describes that the formula \(\varphi\) must hold for all time steps \(t \in
        [a,b]\).
  \item \(\Ev_{[a,b]} \varphi\), where \(a, b \in \Ne \cup \{ \infty \}\),
        describes that the formula \(\varphi\) must hold at least
        once for \(t \in [a, b]\).
\end{itemize}
We refer the readers to \cite{bartocci2018specificationbased} for a detailed
survey of such specification languages.

\begin{table*}
  \caption{Results for motion planning using automata and temporal logic
    objectives with single integrator dynamics.}\label{tab:results-openloop}
  \centering
  \begin{tabularx}{0.95\textwidth}{Xp{2.5cm}ccccccc}
    \toprule
    \multirow{3}{=}{Specification}
                                       & \multirow{3}{=}{Setting}
                                       & \multicolumn{4}{c}{Performance}
                                       & \multirow{2}{*}{\(H\)}
                                       & \multirow{2}{*}{\(k\)}
                                       & \multirow{2}{*}{\(\gamma\)}                                                                                 \\
                                       &                                 & \multicolumn{4}{c}{Num.
    Iterations \(t^*\) (lower better)} &                                 &                                                         &                 \\
                                       &                                 & \multicolumn{4}{c}{Robustness \(\rho\) (higher better)} &       &       & \\
    \cmidrule{3-6}
                                       &
                                       & \((\min,\max)\)                 & \((\max,+)\)                                            & STLCG & MILP
                                       &                                 &                                                         &                 \\
    \midrule
    \multirow{2}{=}{\(\varphi_1\): Visit the RED region, GREEN region and the
      STAR in any order, while avoiding unsafe region, BLUE.}
                                       & Open Loop (\(t^*\))
                                       & 746                             & 78                                                      & 1164  & 42265
                                       & 50                              & 1400                                                    & 0.05            \\
    \cmidrule{2-9}
                                       & Closed Loop (\(\rho\))

    (50 time steps)
                                       & 0.0                             & 0.0                                                     & --    & --
                                       & 15                              & 30                                                      & 0.05            \\
    \midrule
    \multirow{2}{=}{\(\varphi_2\): Visit 3 regions in sequence RED \(\to\)
      GREEN \(\to\) ORANGE at least once while avoiding unsafe region BLUE.}
                                       & Open Loop (\(t^*\))
                                       & 35                              & 425                                                     & 234   & --
                                       & 80                              & 1400                                                    & 0.05            \\
    \cmidrule{2-9}
                                       & Closed Loop (\(\rho\))

    (80 time steps)
                                       & 0.060                           & 0.048                                                   & --    & --
                                       & 40                              & 30                                                      & 0.05            \\
    \midrule
    Adaptive Cruise Control
                                       & Closed Loop (\(\rho\))

    (3000 time steps)
                                       & -1.78                           & 0.4                                                     & -2.21 & 0.6
                                       & 250                             & 30                                                      & 0.05            \\
    \bottomrule
  \end{tabularx}
\end{table*}
%

\begin{figure}
  \begin{center}
    \includegraphics[width=0.45\columnwidth]{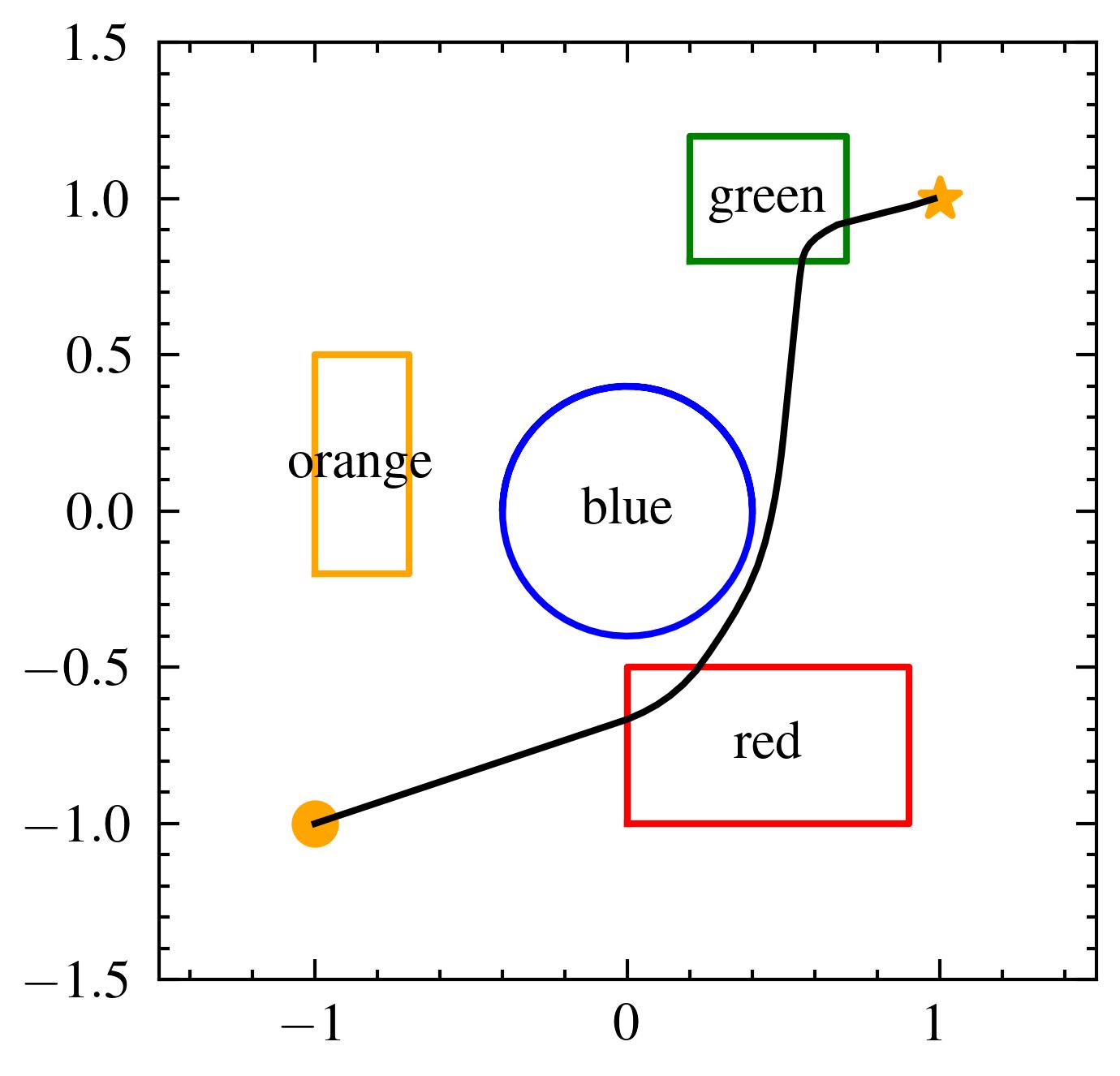}
    \includegraphics[width=0.45\columnwidth]{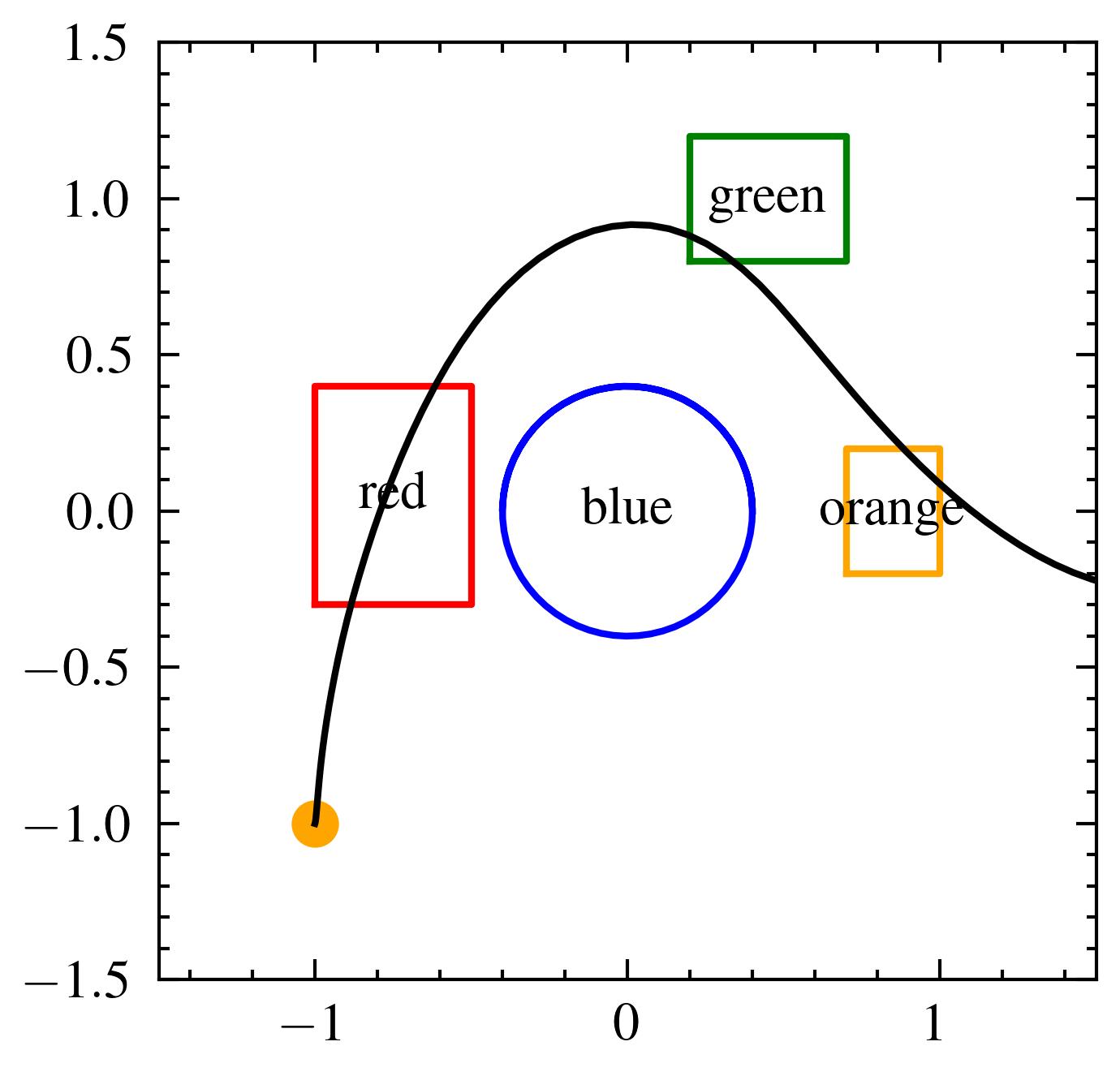}
  \end{center}
  \caption{Example of a trajectory satisfying \(\varphi_1\) and \(\varphi_2\).
    The left-side trajectory was produced by the \((\min,\max)\) automaton in a
    single integrator dynamics model, while the second one was produced by the
    \((\max, +)\) automaton for unicycle model.}\label{fig:phi1_open_loop_eg}
\end{figure}
\paragraph*{\(\varphi_1\): Reach Multiple and Avoid}
Here, we control a point mass on a 2D workspace, governed via simple
single-integrator dynamics by directly controlling its velocity on the
workspace:
\begin{equation}
  x_{t+1} = x_t + u_t \Delta t,
\end{equation}
where the sampling time \(\Delta t = 0.1\) seconds.
The goal of the controller is defined by the specification
\begin{equation}
  \label{eq:phi1}
  \begin{aligned}
    \varphi_1 & := \Ev \Alw_{[0,5]} (x \in \Rc_{\text{red}})      \\
              & \land \Ev \Alw_{[0,5]} (x \in \Rc_{\text{green}}) \\
              & \land \Ev (\abs{x - x_{\text{goal}}} \leq \delta) \\
              & \land \Alw(x \not\in \Rc_{\text{blue}}),
  \end{aligned}
\end{equation}
which says that the controlled mass must visit regions \(\Rc_{red}\) and
\(\Rc_{\text{green}}\) for at least 5 time steps and visit \(x_{\text{goal}}\)
(in any order) while always avoiding \(\Rc_{\text{blue}}\).

In \autoref{fig:phi1_open_loop_eg}, we can see that while all the tested methods
are able to generate accepting trajectories in the open-loop case, the \((\max,
+)\)-automaton matrix operator performs \(\approx 10\) times faster than other
gradient-based methods.

Moreover, in the closed-loop case, only the automaton-based methods seem to be
able to complete the task.
This is because there is no way to encode temporal requirements as above in
these other frameworks, making automata-based methods the only viable option for
long-horizon, temporal tasks.

\paragraph*{\(\varphi_2\): Sequential Tasks with Avoid}
Here, we model the system as a simple unicycle on a 2D workspace,
with each state being represented as \(p = (x, y,
\theta, v, \omega)\): the \(x\)- and \(y\)- positions, the heading, the
linear velocity, and the angular velocity of the unicycle.
The system is thus controlled by the second-order linear and angular
accelerations \(u = (u_a, u_{\omega})\):
\begin{equation}
  \label{eq:unicycle}
  p_{t+1}
  = p_t +
  \begin{bmatrix}
    v \cos(\theta) \\
    v \sin(\theta) \\
    \omega         \\
    u_a            \\
    u_{\omega}
  \end{bmatrix}
  \Delta t,
\end{equation}
where the sampling time \(\Delta t = 0.1\) seconds.
The goal of the controller is to move the unicycle from some initial state to
the regions \(\Rc_{\text{red}}\), \(\Rc_{\text{green}}\), and
\(\Rc_{\text{orange}}\) in that sequence, while avoiding \(\Rc_{\text{blue}}\):
\begin{equation}
  \label{eq:phi2}
  \begin{aligned}
    \varphi_2 & :=                                \\
              & \Ev \of*{
      (x \in \Rc_{\text{red}}) \land
      \Ev \of*{
        (x \in \Rc_{\text{green}}) \land
        \Ev \of*{
          (x \in \Rc_{\text{orange}})
        }
      }
    }                                             \\
              & \land \Alw (x \not\in \Rc_{blue})
  \end{aligned}
\end{equation}

Note that, similar to \(\varphi_1\), this specification cannot be encoded as
a receding horizon controller in STLCG and the MILP encoding, as the history of
what regions have been visited need to be encoded.
Moreover, we do not perform the experiment for the open-loop problem with the
MILP encoding as the system is inherently non-linear, and linearizing it about
discrete \(\theta\) is not a requirement fot the other frameworks.

\paragraph*{Adaptive Cruise Control}

\newcommand{\ego}{\text{ego}}
\newcommand{\lead}{\text{lead}}
\newcommand{\rel}{\text{rel}}

In an adaptive cruise control (ACC) scenario, we are designing a controller for the
trailing car (called \emph{ego} vehicle) such that it maintains a cruising
velocity while also remaining a safe time gap away from the lead car.
Here, the controller operates on the state space
\begin{displaymath}
  x = (p_{\ego}, v_{\ego}, d_{\lead}, v_{\rel}),
\end{displaymath}
where \(p_{\ego}, v_{\ego} \in \Re\) are the longitudinal position and
longitudinal velocity of the ego vehicle; \(d_{\lead} \in \Re\) is the distance
to the lead vehicle; and \(v_{\rel} \in \Re\) is the relative velocity of the
lead car.
While the ego controller receives the actual \(d_{\lead}\) and \(v_{\rel}\) form
an external source, for the MPC prediction step, we (potentially incorrectly)
assume a constant velocity for the lead car.
The control input to the system \(u \in \Re\) is the acceleration of the ego
vehicle, thus we can write an approximate \emph{predictive} model of the system
as:
\begin{equation}
  \label{eq:car-dynamics}
  \begin{split}
    p_{\ego,t+1}  & = p_{\ego,t} + v_{\ego,t} \Delta t + \frac{1}{2} u_t \Delta t^2 \\
    v_{\ego,t+1}  & = v_{\ego,t} + u_t \Delta t                                     \\
    d_{\lead,t+1} & = d_{\lead,t} + v_{\rel,t}\Delta t - \frac{1}{2} u_t \Delta t^2 \\
    v_{\rel,t+1}  & = v_{\rel,t} - u_t \Delta t
  \end{split}
\end{equation}
where the sampling time \(\Delta t = 0.01\) seconds.

\newcommand\VRef{\text{ref}}
\newcommand\Safe{\text{safe}}
\newcommand\Follow{\text{follow}}

The requirements for the behavior of an ACC is parameterized by a target cruise
speed \(v_\VRef\) that the ego must reach if safe to do so; a safety distance
\(d_{\Safe}\) that the ego vehicle must not cross when trailing a car; and a
\emph{safe time gap} \(t_{\Safe}\), which is the time threshold to violating
\(d_\Safe\).
Thus, we can define the requirements by the following:
\begin{align}
  \varphi_{\text{safe}} & = \Alw(d_\lead > d_\Safe) \\
  \varphi_{\VRef}       & =
  \begin{multlined}[t]
    \Alw((d_\lead > d_\Follow)\\
    \quad\Rightarrow \Ev_{[0,\delta]}
    (\abs{v_\lead - v_\VRef} < \epsilon \lor d_\lead \leq d_\Follow)),
  \end{multlined}
\end{align}
where \(d_\Follow = d_\Safe + v_\rel t_\Safe\) is a safe following distance that
doesn't violate the safe time gap, with the parameters \(v_\VRef = 15
\text{m}/\text{s}\), \(t_\Safe = 1.4\text{s}\), \(d_\Safe = 5\text{m}\), \(\delta = 50\) time steps and \(\epsilon
= 1 \text{m}/\text{s}\).
The above specification \(\varphi_\VRef\) describes that the controller must
reach the target speed if the safe time gap isn't violated.

In this scenario, we can see from \autoref{tab:results-openloop} that the
gradient-based methods that use \((\min,\max)\)-based semantics seem to get
stuck in local optima, similar to the analysis in~\cite{pant2017smooth}.
This isn't a problem for the \((\max,+)\) semiring as replacing \(\min\)
operations with standard addition makes the problem more conducive
to gradient-based optimization, while preserving acceptance semantics.


\section{Conclusion}

In this paper, we target encoding automata-based specifications into
quantitative objective functions, allowing efficient encoding of signal history,
along with quantitative cost functions for motion planning.
Specifically, we define a automaton matrix operator that encodes transitions in
the automaton, thereby leveraging matrix-based automatic differentiation tools
for gradient-based motion planning.
To the best of our knowledge, this is the first such framework that combines
optimization-based and automata-based methods to leverage modern compute
capabilities to directly optimize over automaton specifications efficiently.
Moreover, to mitigate issues with local optima and vanishing gradients in
the quantitative semantics, we show how using an alternative
\((\max,+)\)-algebraic semantics (as presented in \cite{jaksic2018algebraic})
allows us to find satisfying trajectories faster.


\bibliographystyle{IEEEtran}
\bibliography{main}
\end{document}